\documentclass[runningheads]{llncs}
\usepackage{graphicx}
\usepackage{tikz}
\usepackage{comment}
\usepackage{amsmath,amssymb} % define this before the line numbering.
\usepackage{color}
\usepackage{CJKutf8}

\usepackage{url}

\newcommand{\chinese}[1]{\begin{CJK}{UTF8}{bsmi}#1\end{CJK}}
\newcommand{\fifth}{\chinese{五} }

\usepackage[accsupp]{axessibility}  % Improves PDF readability for those with disabilities.

\begin{document}
%Hello World
%\begin{comment}
% \linenumbers
\pagestyle{headings}
\mainmatter
\def\ECCVSubNumber{13}  % Insert your submission number here

\title{COVID Detection and Severity Prediction with 3D-ConvNeXt and Custom Pretrainings}
% \title{Custom Pretrainings and Adapted 3D-ConvNeXt Architecture for COVID Detection and Severity Prediction} % Replace with your title
%\title{Sophisticated}


% Platz für Titel reicht nicht für Detection. Was tun?
\titlerunning{ConvNeXt for COVID Detection and Severity Prediction}
%\titlerunning{ECCV-22 submission ID \ECCVSubNumber} 
%\authorrunning{ECCV-22 submission ID \ECCVSubNumber} 
%\author{Anonymous ECCV submission}
%\institute{Paper ID \ECCVSubNumber}

\author{Daniel Kienzle\textsuperscript{*}, Julian Lorenz\textsuperscript{*}, Robin Schön\textsuperscript{*}, Katja Ludwig, Rainer Lienhart}

\authorrunning{D. Kienzle, J. Lorenz, R. Schön et al.}

\institute{
Augsburg University, Augsburg 86159, Germany \\
%ANONYMOUS \\
\email{\{firstname.lastname\}@uni-a.de}
%\email{\{firstname.lastname\}@ANONYMOUS}
}

%\institute{Princeton University, Princeton NJ 08544, USA \and
%Springer Heidelberg, Tiergartenstr. 17, 69121 Heidelberg, Germany

%******************
\maketitle
\begingroup\renewcommand\thefootnote{$*$}
\footnotetext{Authors contributed equally}
\endgroup

\begin{abstract}
Since COVID strongly affects the respiratory system, lung CT-scans can be used for the analysis of a patients health. We introduce a neural network for the prediction of the severity of lung damage and the detection of a COVID-infection using three-dimensional CT-data. Therefore, we adapt the recent ConvNeXt model to process three-dimensional data. Furthermore, we design and analyze different pretraining methods specifically designed to improve the models ability to handle three-dimensional CT-data. We rank 2nd in the \emph{1st COVID19 Severity Detection Challenge} and 3rd in the \emph{2nd COVID19 Detection Challenge}.

\keywords{Machine Learning, COVID Detection, Severity Prediction, Medical Image Analysis, CT scans, 3D data}
\end{abstract}

\section{Introduction}
The last few years have been strongly shaped by the COVID-19 pandemic, with a considerable amount of cases ending deadly. For the treatment of patients it is crucial to predict the severity of lung damage caused by a SARS-CoV-2 infection accurately. The lung damage is visually detectable by visible ground-glass opacities and mucoid impactions on the slices of a patients CT-scan (\cite{stoic}). Thus, it might be beneficial to automatically process CT-scans for the diagnosis of the patients.

In this paper, we introduce a neural network to automatically analyze CT-scans. We train our model to classify the severity of lung damage caused by SARS-CoV-2 into four different categories. The model is trained and evaluated using the COV19-CT-DB database (\cite{kollias2021mia}). Additionally, we transfer our architecture and training pipeline to the detection of SARS-CoV-2 infections in CT-scans and train a separate model for this task. Consequently, we show that our method can easily be transferred to multiple COVID-related analyses of CT-scans. We rank 2nd in the \emph{1st COVID19 Severity Detection Challenge} and 3rd in the \emph{2nd COVID19 Detection Challenge}. Moreover, our model is especially good at identifying the most severe cases that are most important to detect in a clinical setting.

As medical datasets are small in comparison to common computer-vision datasets, the application of large computer-vision architectures is not straight forward as they tend to overfit very quickly. As a result, the development of a good pretraining pipeline as well as the utilization of additional data is essential in order to get adequate results.

Since medical datasets are comparably small, the validation split is as well in most cases very small. However, evaluating the models performance on a single small validation set leads to non-representative results as the validation set is not representative for the overall data distribution. Furthermore, the evaluation on a single small dataset could cause overfitting of the hyperparameters to the validation set characteristics and, therefore, reduces the models test-set performance. As a result, it is very important to use strategies like cross-validation in order to get a better estimate of the models performance.

Goal of this paper is to develop a neural network that is capable of automatically predicting four degrees of severity of lung damage from a patents lung CT-scan. In addition, we also adapt our architecture to predict infections with the SARS-CoV-2 virus using CT-scans. In order to improve the performance on these two tasks, our main contributions are:
\begin{enumerate}
    \item We adapt the recent ConvNeXt architecture (\cite{liu2022convnet}) to process three dimensional input-data. 
    \item We introduce multiple techniques for pretraining of our architecture in order to increase the ability of our network to handle three-dimensional CT-scans. 
\end{enumerate}

\section{Related Work}
% Neural networks for medical predictions 
The idea of using neural networks for the prediction of certain properties visible in medical data has developed to increasing levels of importance in the last few years (examples can be found in \cite{suganyadevi2022review}, \cite{abdou2022literature}, \cite{wang2021review}). The authors of \cite{kollias2018deep}, \cite{kollias2020deep}, \cite{arsenos2022large} and \cite{kollias2020transparent} have used CNNs and Recurrent Neural Networks (RNNs) for the prediction of Parkinson's disease on brain MRI and DaT scans. In \cite{7918014}, \cite{shakeel2020automatic}, \cite{9128479} NN-based methods for the detection of lung cancer are developed.

% Usage of neural networks for the detection of COVID-19 
Since its occurrence in late 2019, a considerable number of articles have concerned themselves with using neural networks for the purpose of predicting a potential SARS-CoV-2 infection from visual data. 
% Purely neural
The authors of \cite{mukherjee2021deep} propose neural networks for the usage of CT scans as well as chest x-rays, whereas \cite{POLSINELLI202095} puts a lot of focus on computational efficiency and design a lightweight network, in order to be able to also run on CPU hardware. 
% Mixed method
In the wake of this development, there have also been methods which combine neural and non-neural components. In \cite{BASU2022116377} the authors first extract the features by means of neural network backbone, and the utilize an optimization algorithm in combination with a local search method before feeding the resulting features into a classifier. 
The method of \cite{kundu2021fuzzy} applies a fusion based ranking model, based on reparameterized Gompertz function, after the neural network has already produced its output probabilities. 
In \cite{kollias2020transparent} and \cite{kollias2020deep} the authors also make use of clustering in order to carry out a further analysis of the produced features vectors, and classify the CT scans according to their proximity to the cluster centers. 

% ICCV 2021 
In the context of last years ICCV there has been a challenge with the aim of detecting COVID-19 from CT images (\cite{kollias2021mia}). The winners of this contest (\cite{Hou_2021_ICCV}) were using contrastive learning techniques in order to improve their networks performance. \cite{Miron_2021_ICCV} and \cite{Liang_2021_ICCV} use 3D-CNNs for the detection of the disease, whereas the teams in \cite{Zhang_2021_ICCV} and \cite{Tan_2021_ICCV} happen to use transformer-based architectures. 
In addition to that, \cite{Anwar_2021_ICCV} proposes the usage of AutoGluon (\cite{AutoGluon}) as an AutoML based approach.

% ConvNext architecture and transformer based methods 
Our architectures is, similar to other previously existing approaches for the processing of 3D-data, based on the idea, that 2D architectures can be directly extended to 3D architectures (\cite{i3d_architecture}, \cite{ruiz20203d}, \cite{kopuklu2019resource}). In our case we use 3D modification of the ConvNeXt architecture (\cite{liu2022convnet}). This particular approach is characterized by architectural similarities to MobileNets (\cite{howard2017mobilenets}, \cite{Sandler_2018_CVPR}) and Vision Transformers Transformers (ViT, \cite{visiontransformer}, \cite{liu2021Swin}). 

% Pretraining semi-supervision
In the medical field, due to privacy restrictions that protect the patients' data, the potential amount of training data is rather sparse. This especially holds, when it comes to datasets that accompany particular benchmarks. However, datasets for similar tasks may be exploited for pretraining and multitask learning, if one can assume that insights from one datasets might benefit the main objective. This inspired the authors of \cite{tang2022self} and \cite{hatamizadeh2022swin} to pretrain the model on pretext tasks, which are carried out on medical data. The authors of \cite{google_pretraining} show that pretraining on ImageNet is useful, by the virtue of the sheer amount of data. 
In some particular cases, we might have access to a larger amount of data while lacking labels. The publications \cite{seibold2022reference} and \cite{chaitanya2021contrastive} use semi-supervised learning techniques to overcome this particular situation.

% All of them are done, apart from cross validation, for which it would be unusual to cite papers, since it 
% is relatively common practice. 
%\todo{This is the old related work. Where there important changes that are not included yet?}
%\todo{pretraining; semi-supervision; cross-validation(?)}
%\todo{Paper vom Chef zitieren}

\section{Methods}
Goal of this work is to develop a neural network architecture capable of predicting the severity of a SARS-CoV-2 infection and to transfer the method to the task of infection detection. We apply our models to the COV19-CT-DB database (\cite{kollias2021mia}). The train and test set of this database consist of 2476 scans for the task of infection detection and 319 images for severity prediction. Each scan is composed of multiple two-dimensional image slices (166 slices on average). We concatenate these slices into a three-dimensional tensor and apply cubic spline interpolation to get tensors of the desired spatial dimension. In this section we introduce the key methods for improving the automatic analysis of this data.

\subsection{ConvNeXt 3D}
The architecture we utilize is a three-dimensional version of the recent ConvNeXt architecture \cite{liu2022convnet}. This architecture type is especially characterized by multiple alterations, which have already proven themselves to be useful in the context of Vision Transformers, and were applied to the standard ResNet (\cite{He_2016_CVPR}). 

% What is ConvNeXt
For example, ConvNeXt has a network stem that patchifies the image using non-overlapping convolutions followed by a number of blocks with a compute ratio of $(3:3:9:3)$ that make up the stages of the network. The influence which MobileNetV2 \cite{Sandler_2018_CVPR} had on this type of architecture is expressed by the introduction of inverted bottleneck blocks and the usage of depthwise convolutions, which, due to their computational efficiency, allow for an unusually large kernel size of $7 \times 7$. Additional distinguishing properties of this architecture are the replacement of Batch Normalization by Layer Normalization, the usage of less activation functions and the replacement of the ReLU activation function by the GELU activation \cite{geluactivation}.

% The extension to the 3D model (especially the inflation of the weights (!!!)). 
The standard ConvNeXt architecture in its initial form was conceptualized for the purpose of processing 2D images with 3 color channels, whereas we want to process 3D computational tomography scans that only have one color channel (initially expressed in Hounsfield Units \cite{buzug2011einfuhrung}). We adapted the ConvNeXt architecture to our objective by using 3D instead of 2D convolutions. In order to be able to make use of potentially pre-existing network weights, we apply kernel weight inflation techniques to the 2D networks parameters as described in section \ref{section:pretraining}.

\subsection{Pretraining}
\label{section:pretraining}

In contrast to ordinary computer vision datasets like e.g. ImageNet \cite{Imagenet}, medical datasets are usually considerably smaller. In order to still be able to train large neural networks with these datasets we utilize various pretraining techniques. As our data consists of three-dimensional gray-scale tensors as input data instead of two-dimensional RGB-images and we use 3D-convolutions instead of 2D-convolutions, it is not possibly to directly use the publicy-available pretrained ConvNeXt weights. In this section we present various possibilities for the initialization of our network with pretrained weights.

For 2D models, it is common to pretrain a model on the task of ImageNet classification. As our data consists of gray-scale tensors, we implemented a ConvNeXt pretraining with gray-scale ImageNet images to obtain weights for a two-dimensional ConvNeXt model. 
To use those weights for our 3D model, we propose three different inflation techniques for the two-dimensional weights of the pretrained 2D model. We will refer to these as \emph{full inflation}, \emph{1G inflation} and \emph{2G inflation}. 

Let $\mathtt{K} \in \mathbb{R}^{I \times O \times H \times W}$ be the 2D kernel weight tensor, and $\mathtt{K}^\uparrow \in \mathbb{R}^{I \times O \times H \times W \times D}$ be the 3D kernel weights after inflation. For these kernel weights we denote by $I$ the input channels, $O$ the output channels, $H$ the height, $W$ the width and $D$ the additional dimension of the 3D kernel. Also, let $i, o, h, w, d$ denote all possible positions along the aforementioned dimensions. $\gamma$ is a normalization factor that normalizes the inflated tensor $\mathtt{K}^\uparrow $ to have the L2 norm of the 2D kernel $\mathtt{K}$.

 The first way, called \emph{full inflation}, is the commonly used option of simply copying the weights along the new tensor axis  \cite{i3d_architecture}. 
This can be described as an equation of the form 
\begin{equation}
\forall i, o, h, w, d \; : \; \mathtt{K}^\uparrow_{i, o, h, w, d} = \mathtt{K}_{i, o, h, w} \cdot \gamma.
\end{equation}
In \emph{1G inflation}, we use a Gaussian weight $\mathcal{N}\left(\cdot, \mu, \sigma\right)$ in order to create different weights, that are the largest in the kernel center: 
\begin{equation}
\forall i, o, h, w, d \; : \; \mathtt{K}^\uparrow_{i, o, h, w, d} = \left( \mathtt{K}_{i, o, h, w} \cdot \mathcal{N}\left(d, \frac{D}{2}, \frac{D}{8}\right) \right) \cdot \gamma
\end{equation}

\begin{figure}
    \centering
    \begin{tabular}{ccc}
         \includegraphics[width=70pt]{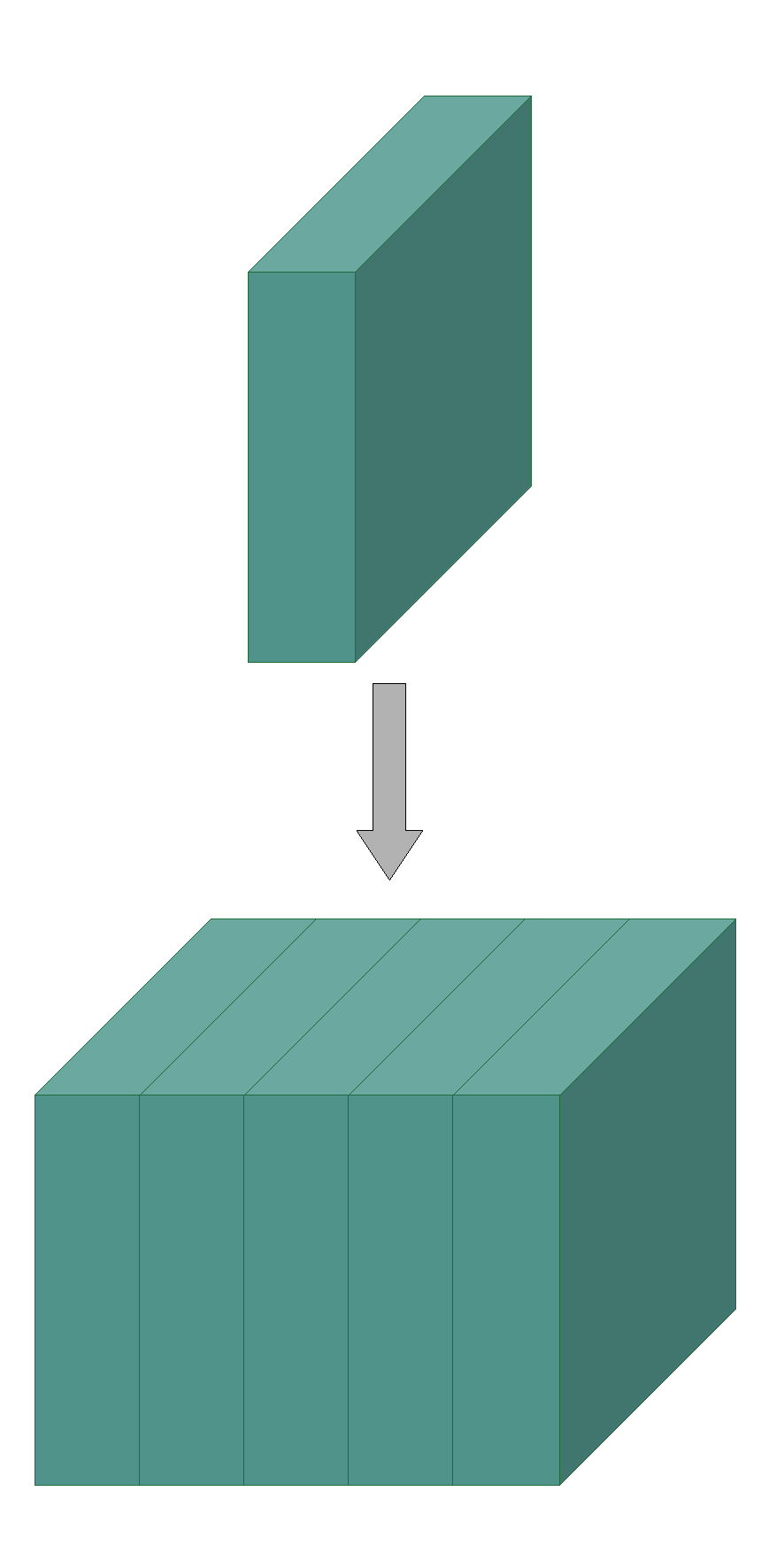} & \includegraphics[width=70pt]{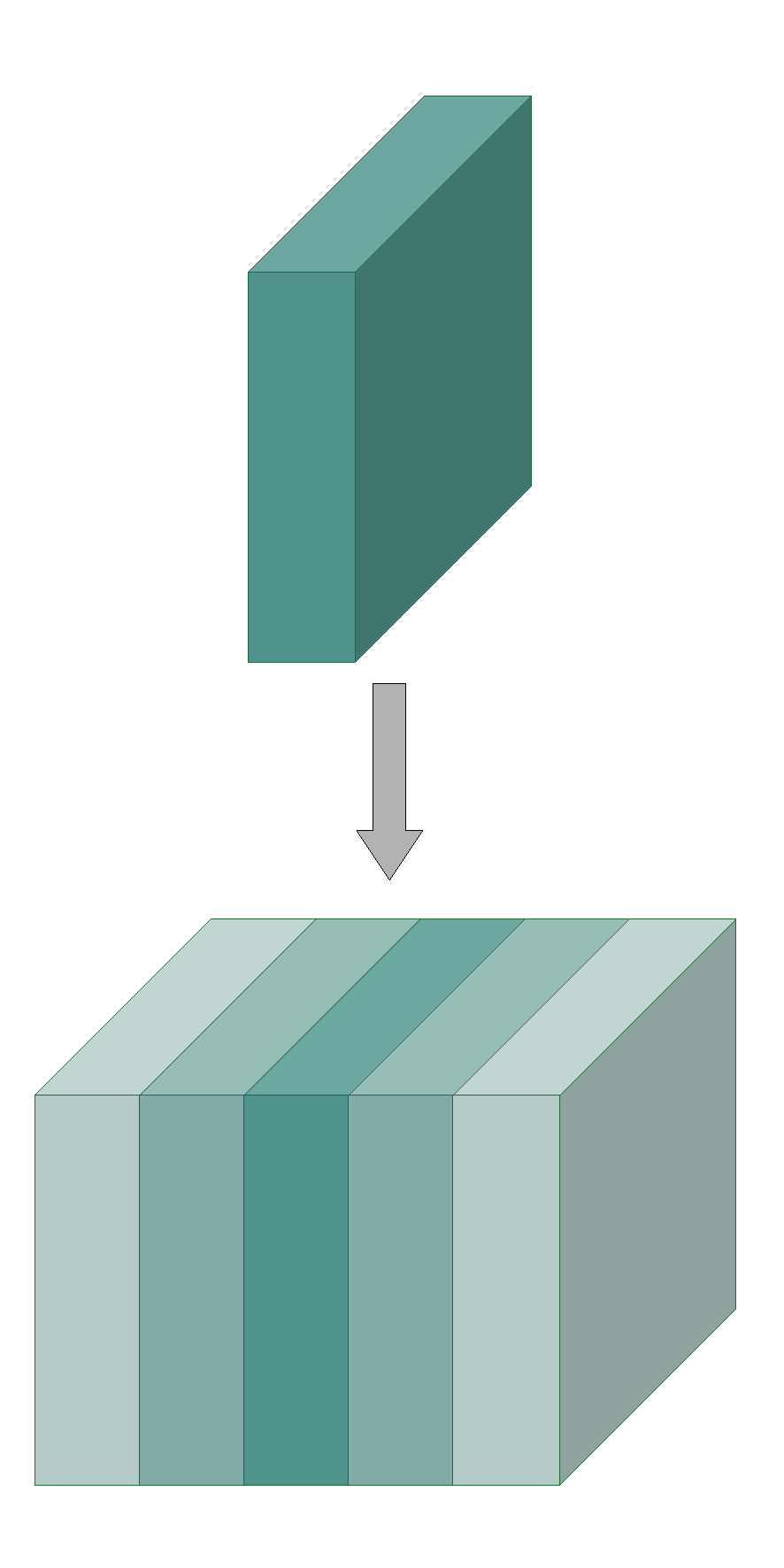} & \includegraphics[width=140pt]{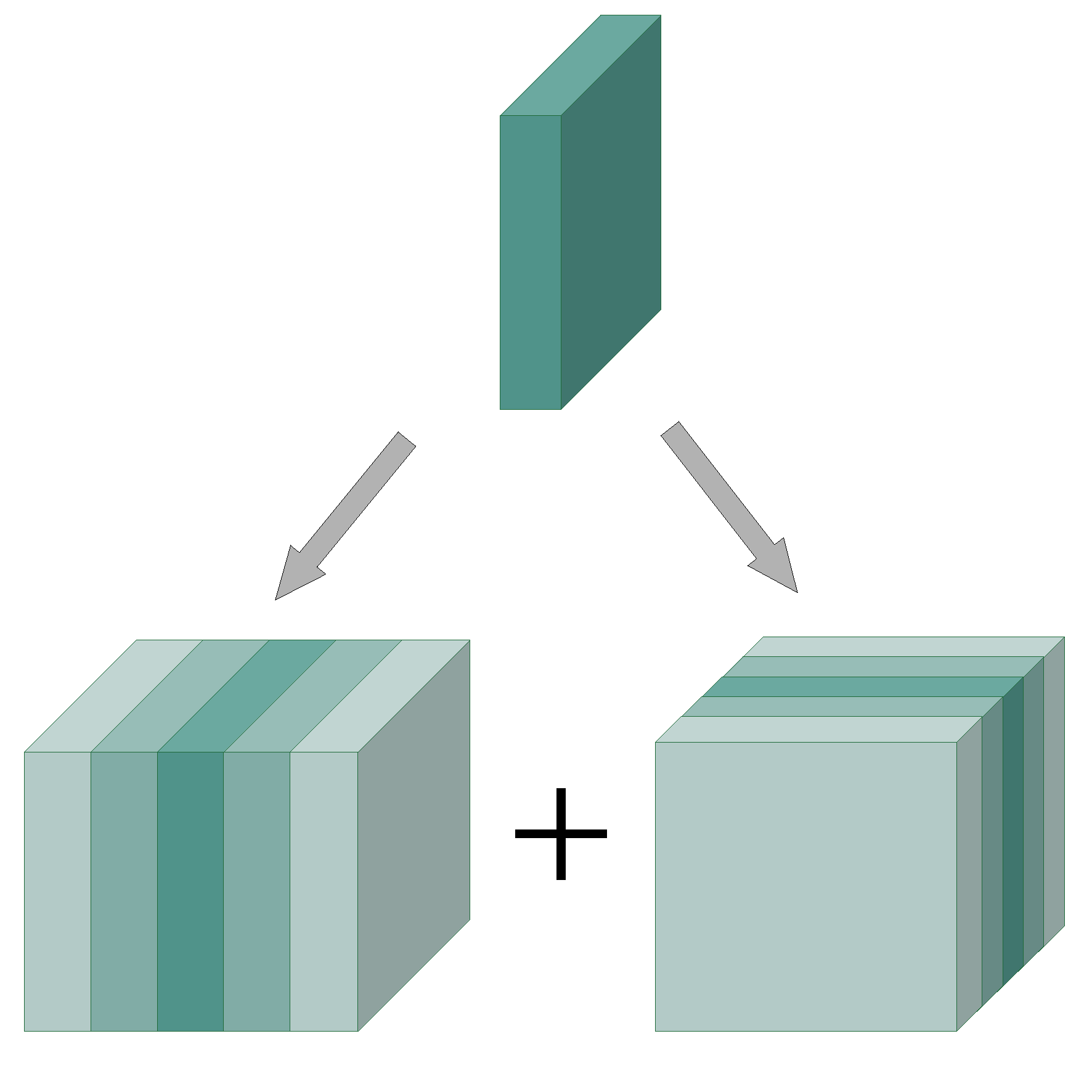} \\
         Full Inflation & 1G Inflation & 2G Inflation 
    \end{tabular}
    \caption{A visualization of the different inflation approaches to generate 3D kernels. \emph{Full inflation} simply copies the weights along the new axis. \emph{1G inflation} also copies the weights along the new axis but multiplies them with Gaussian weights. \emph{2G inflation} acts similar to 1G inflation but add the weights after going over two dimensions. }
    \label{fig:inflation}
\end{figure}

The third, that is referred to \emph{2G inflation} is based on multiplying the 2D weights along 2 axes:  

\begin{equation}
\begin{split}
\forall i, o, h, w, d \; &: \\
\; \mathtt{K}^\uparrow_{i, o, h, w, d} &= \left( \mathtt{K}_{i, o, h, w} \cdot \mathcal{N}\left(d, \frac{D}{2}, \frac{D}{8}\right) + \mathtt{K}_{i, o, h, w} \cdot \mathcal{N}\left(w, \frac{W}{2}, \frac{W}{8}\right) \right) \cdot \gamma
\end{split}
\end{equation}  
The different inflation approaches to create 3D kernels are visualized in Figure \ref{fig:inflation}. \\

Since the images in the ImageNet database are very distinct from CT-images as used in this paper, we introduce various further ways to adjust the model to three-dimensional CT-scans. For instance, we use an additional dataset designed for lung-lesion segmentation in CT scans (\cite{roth2021rapid}, \cite{tcia_data_2}, \cite{clark2013cancer}) and the STOIC dataset created for SARS-CoV-2 severity prediction (\cite{stoic}). As those datasets consist of CT-scans of SARS-CoV-19 infected patients similar to the COV19-CT-DB database, we assume that pretraining with these additional datasets will be beneficial for our model performance and will increase in robustness as it is able to deal with a greater variety of data.

Since the lung damage caused by SARS-CoV-2 is visually detectable in lung CT-scans, we use the segmentation dataset to pretrain our model to segment lung lesions. By directly showing the damaged lung regions to the network we hope to provide a reasonable bias for learning to predict the severity. Furthermore, a segmentation pretraining is beneficial as the segmentation task is more robust to overfitting in contrast to a classification task. This is important as it enables us to apply large-scale architectures to the small medical datasets.

The STOIC dataset provides two categories of severity for each patients CT-scans. Even though the categories in the STOIC dataset are different to the categories in the COV19-CT-DB database we assume, nevertheless, that pretraining with the STOIC dataset teaches the network a general understanding of severity.

\begin{figure}
    \centering
         \includegraphics[width=280pt]{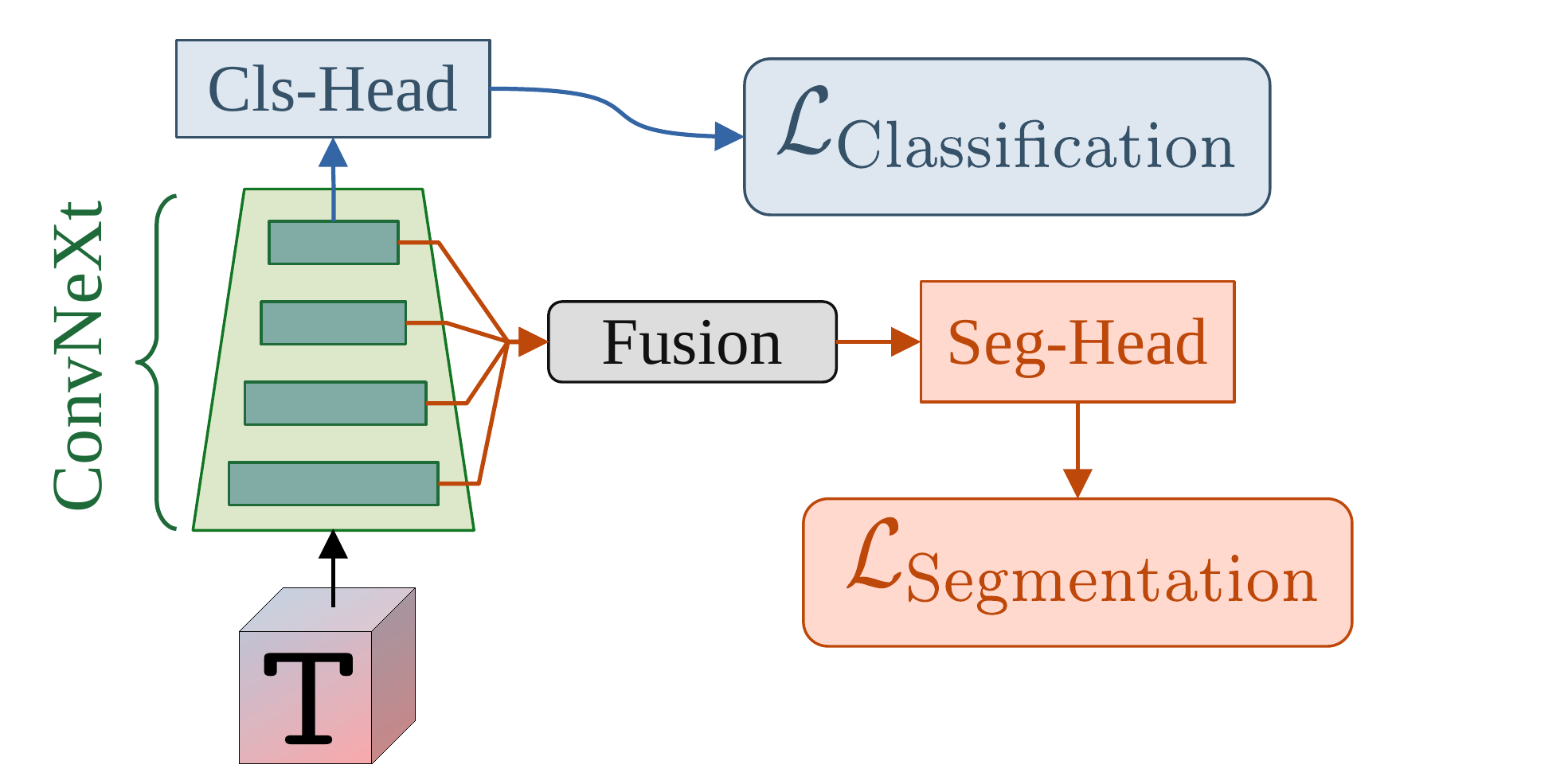}
    \caption{Visualization of pretraining architecture. An \emph{input-tensor} $\mathtt{T}$ is processed by the \emph{ConvNeXt} model. To generate a classification label (e.g. for severity prediction or infection detection) the output-features of the last block are processed by a task-specific \emph{classification head} in order to generate the class probabilities. In order to compute a segmentation mask, the output-features of every block of the ConvNeXt architecture are upsampled and concatenated similar to \cite{upernet}. This is further processed by a \emph{segmentation-head} in order to produce the segmentation output. }
    \label{fig:multinext}
\end{figure}

In order to be able to generate segmentation masks with our model additionally to severity classification outputs, we extend our architecture similar to the Upernet architecture \cite{upernet}. Our architecture is explained in figure \ref{fig:multinext} 
 \\
% Here are the different second stage pretrainings 
In this work we compare 4 different pretraining methods:
\begin{enumerate}
    \item We directly use our inflated grayscale ImageNet pretraining weights. This approach is referred to as \emph{ImageNet model.}
    \item We train our network for the task of segmentation on the segmentation dataset starting from inflated ImageNet weights (1.). This approach is referred to as \emph{segmentation model}.
    \item Pseudo-labels are generated with the segmentation model (2.) in order to get segmentation masks for the COV19-CT-DB database. We train a new model for the task of segmentation on the COV19-CT-DB database using the pseudo-labels as ground-truth. The model is initialized with our inflated ImageNet weights (1.). This is referred to as \emph{segmia model}.
    \item Pseudo-labels are generated with the segmentation model (2.) to get segmentation masks for the STOIC dataset. We train a new model to jointly optimize severity classification using real labels and segmentation using pseudo-labels for the STOIC dataset. This model is initialized with our inflated ImageNet weights (1.). This is referred to as \emph{multitask model}
\end{enumerate}
After each pretraining method, we finetune our model for either severity prediction or infection detection on the COV19-CT-DB database.

\subsection{Approaches for Increased Robustness}
\label{increasedrobustness}
%\todo{This is the old section. Things from the arxiv paper still have to be added!}
%TODO: \\
%explain Augmentations. Especially talk about our elastic deformation
%\todo{Motivation für Augmentierung}
% I'm not exactly sure, as to what was meant by "motivation", but this might be it: 
When automatically analyzing CT images, we have to account for multiple potential forms of irregularities. Not only might different CT scanners result in varying image details, but the person in the scanner might also lie differently on every image. We thus have to increase the models robustness. 

We use the following classical augmentations: Random flips along all three axes, Gaussian noise with random standard deviation between 0.6 and 0.8, and Gaussian blur. Since some patients lie in the CT scanner with a slight inclination to the side of their body, we rotate the tensors along the transversal axis by a randomly picked angle from the interval $(-30^\circ, 30^\circ)$. 

As is common practice in the field of medical computer-vision, we also use elastic deformations (with a chance of $50\%$). For the generation of those deformations the vector field is scaled by a randomly drawn $\alpha \in (1, 7)$ and then smoothed with a Gaussian kernel with $\sigma = 35$. We create our own GPU compatible implementation, and decompose the used Gaussian filter along its axes. This results in an augmented computational speed, rendering the deformations viable for fast online computation. 

During manual inspection, we came across some tensors that were oriented in a different way (for example vertically instead of horizontally) or with patients facing a different direction. In order to make our network robust to these variations, we add an augmentation that simulates different orientations: For each of the three axes ($x$-, $y$- and $z$-axis) we randomly pick a multiple of $90^\circ$, and rotate the tensor by this angle. Since the missoriented tensors still constitute a considerable minority of all cases we do not apply this augmentation to every tensor during training, but only with a probability of 25\%.

%From now on we will refer to models trained with this augmentation as ROr. 

%On elastic deformation: Self Made for gpu (paralellization very fast) 

%\subsection{Data preprocessing}
%TODO: \\
%interpolation of 2D slices to 3D tensor; resolution; precomputing two different resolutions (before and after crop) in order to reduce computation time for downsampling; filter out slices with wrong resolution

In addition, we apply random crops with a probability of 50\%. Therefore, we first rescale the scan to a resolution of $(256  \times 256  \times 256)$ and take a random crop of $(224  \times 224  \times 224)$. When no random crop is applied, we directly rescale to the latter resolution. Rescaling of the tensors is performed with cubic spline interpolation and the rescaled tensors are precomputed prior to the training to decrease computation time.  
When inspecting the data, we discovered some tensors where the slice resolution happened to be internally inconsistent between different slices of the same CT. In those cases, we discarded the inconsistent slices. 
In order to stabilize our performance, we kept a second copy of our model whose weights were not learned directly, but which is an exponential moving average (EMA) of the trained models weights. This copy of the network is used for the evaluations and final predictions.

Besides data augmentation, we also use 5-fold cross validation to improve model robustness. We split the public training set into five folds with almost equal size and make sure that each class is evenly distributed across all folds. For example, each fold for severity prediction contains 12 or 13 moderate cases. Each fold forms the validation set for one model, the remaining folds serve as the training set. This way we get 5 models that are trained and evaluated on different datasets. To get the final predictions for the official test set, we predict every case from the test set using all of the 5 models and take the mean of the model outputs. Before averaging the outputs, we apply Softmax to bring all values to the same scale.

\section{Experiments}
Most experiments were performed for the task of severity prediction because of two reasons. First, it is supposedly the more challenging task due to the classification into four classes and, second, it is computationally less expensive due to a smaller dataset allowing for a greater number of experiments.

We evaluate our models using the COV19-CT-DB database. As the validation set is comparatively small, evaluating the models' performance on its validation set leads to non-representative results as the validation set is not representative for the data. Furthermore, evaluation on a single validation set easily causes overfitting of the hyperparameters. As a result, we do not use the validation set to analyse the models performance and instead perform 5-fold cross-validation only on the training set. The average performance of the 5 training runs is used as an estimate for the performance. In order to generate predictions for the validation set and test set we use an ensemble consisting of the 5 models from the cross-validation runs. The predictions of the models are averaged after application of the softmax to produce the final prediction. In addition, we also add the results obtained only with the 5th model.
\subsection{Preliminary Experiments}
As the frequency of the classes in the COV19-CT-DB database is not evenly distributed, we recognized that our neural network trained with ordinary cross-entropy as loss function has problems detecting the less frequent classes. Especially patients with critical severity are not correctly classified. As it is especially important to recognize the critical cases, we introduced a weight in order to balance the cross entropy. Thus, the loss for every class is multiplied with the normalized class-frequency. The severity-prediction results for the ordinary cross entropy (\emph{CE}) and balanced cross entropy (\emph{balanced CE}) are given in table \ref{table:lossFunctions}. The cross-validation performance of both loss functions is very similar. Even though the ordinary cross-entropy performs a little bit better, we chose to use the balanced cross entropy for our further experiment since we think that the balanced cross entropy improves the performance for the underrepresented critical cases. We assume that the good performance of our challenge submission (see table \ref{table:comparisonSeverity}) for critical cases is partly because of the balanced cross-entropy. Consequently, we would advise to use balanced cross-entropy instead of ordinary cross-entropy in a clinical setting. \\
\begin{table}[ht]
\caption{Comparison of balanced cross entropy with ordinary cross entropy. The models are initialized with full ImageNet initialization. The cross validation results and the results for the official validation set are reported. The ensemble predictions are marked with a \textdagger . F1 scores are macro F1 scores}
\begin{center}
\footnotesize
\begin{tabular}{ c| p{18mm} p{16mm} p{14mm} p{18mm} p{14mm} p{14mm}} 
\hline
 \textbf{loss} & \textbf{F1 Cross \newline Val} & \textbf{F1 Val} & \textbf{F1 Val \newline Mild} & \textbf{F1 Val \newline Moderate} & \textbf{F1 Val \newline Severe} & \textbf{F1 Val \newline Critical}\\ 
 \hline
 % compareLossfnCV_balce-imagenet_lrDecayUntil0_initModefull_rotProb0_AugModeall_cv_convnext_20220706-135516
 balanced CE & {64.99} & 62.82\textsuperscript{\textdagger} & 82.93\textsuperscript{\textdagger} & 57.14\textsuperscript{\textdagger} & 57.89\textsuperscript{\textdagger} & 53.33\textsuperscript{\textdagger}  \\ 
 % compareLossfnCV_ce-imagenet_lrDecayUntil0_initModefull_rotProb0_AugModeall_cv_convnext_20220707-140856
 CE & \textbf{65.37} & 60.10\textsuperscript{\textdagger} & 82.05\textsuperscript{\textdagger} & 50.00\textsuperscript{\textdagger} & 55.00\textsuperscript{\textdagger} & 53.33\textsuperscript{\textdagger}  \\ 
 \hline
\end{tabular}
\end{center}
\label{table:lossFunctions}
\end{table}

\subsection{Comparison of Pretrainings}
One main goal of this paper is to enhance the severity-prediction performance by introducing several pretrainings as explained in section \ref{section:pretraining}.
Results for the various ImageNet inflation methods are added in table \ref{table:imagenet}.
\begin{table}[ht]
\caption{Comparison of ImageNet initialization. The cross validation results and the results for the official validation set are reported. The ensemble predictions are marked with a \textdagger . F1 scores are macro F1 scores}
\begin{center}
\footnotesize
\begin{tabular}{ c| c c} 
\hline
 \textbf{initialization} & \textbf{F1 cross validation} & \textbf{F1 validation}  \\ 
 \hline
 % FinalSeverityCV_balce-imagenet_lrDecayUntil0_initModefull_rotProb0.0_AugModeall_cv_convnext_20220621-115412
 full & 65.67 & \textbf{ 61.28\textsuperscript{\textdagger}}  \\ 
 % FinalSeverityCV_balce-imagenet_lrDecayUntil0_initModeone_g_rotProb0.0_AugModeall_cv_convnext_20220706-150338
 1G & 65.81 & 60.71 59.93\textsuperscript{\textdagger}  \\ 
 % FinalSeverityCV_balce-imagenet_lrDecayUntil0_initModetwo_g_rotProb0.0_AugModeall_cv_convnext_20220630-224731
 2G & \textbf{69.61} &  56.92\textsuperscript{\textdagger} \\
 \hline
\end{tabular}
\end{center}
\label{table:imagenet}
\end{table}
According to these results, the performance is best for 2G inflation in terms of the cross-validation metrics and, consequently, we assume that using multiple geometrical-oriented planes is beneficial as it better utilizes all three dimensions. \\ 
Results for the various pretraining methods can be seen in table \ref{table:pretrainings}.
\begin{table}[htb]
\caption{Comparison of models initialized with different pretrainings. Performance is evaluated for the severity-prediction task. Random-initialization is denoted as \emph{Random}. The ensemble predictions are marked with a \textdagger . F1 scores are macro F1 scores.}
\begin{center}
\begin{tabular}{c | c | c | c} 
\hline \textbf{Pretraining} & \textbf{F1 Cross Validation} & \textbf{F1 Test} & \textbf{F1 Validation} \\ 
\hline
% FinalSeverityCV_balce-random_lrDecayUntil0_initModefull_rotProb0.0_AugModeall_cv_convnext_20220711-110251
Random & 62.71 & - &  56.46\textsuperscript{\textdagger} \\
% FinalSeverityCV_balce-imagenet_lrDecayUntil0_initModefull_rotProb0.0_AugModeall_cv_convnext_20220621-115412
ImageNet (Full) & 65.67 &  45.73\textsuperscript{\textdagger} & 61.28\textsuperscript{\textdagger} \\
% FinalSeverityCV_balce-segmentationECCVFull_lrDecayUntil0_initModefull_rotProb0.0_AugModeall_cv_convnext_20220624-183130
Segmentation & 67.25 & 46.21\textsuperscript{\textdagger} & 61.28\textsuperscript{\textdagger} \\
% FinalSeverityCV_balce-segmiaECCVFull_lrDecayUntil0_initModefull_rotProb0.0_AugModeall_cv_convnext_20220625-180946
Segmia & 66.48 & 48.85\textsuperscript{\textdagger} & 63.05\textsuperscript{\textdagger} \\
% FinalSeverityCV_balce-multitaskECCV_lrDecayUntil0_initModefull_rotProb0.0_AugModeall_cv_convnext_20220621-115510
Multitask & \textbf{68.18} & \textbf{48.95\textsuperscript{\textdagger}} & 58.77\textsuperscript{\textdagger} \\
\hline
\end{tabular}  
\end{center}
\label{table:pretrainings}
\end{table}
For some of those experiments, the scores for the official test set of the COV19-CT-DB database are available. As this test set is substantial larger than the official validation set, the performance estimate is considered as more accurate. Thus, we use this metric in addition to our cross-validation results to interpret our performance. The results clearly indicate that the cross-validation metrics is a much better estimate of the models performance than the validation-set metrics since the \emph{segmentation}, \emph{segmia} and \emph{multitask model} perform better than the \emph{ImageNet model} on both cross-validation metrics and test-set metrics. Moreover, the best model in terms of cross-validation metrics performs also best on the test-set. As a result, we strongly advise to evaluate models intended for clinical usage based on cross-validation metrics.

Even though cross validation gives well reasoned clues about the models performance, a large gap between the cross-validation score and the test-set metrics can be observed. Because the test-set performance is worse by a large margin, we suppose that the test-set statistics do not fully match the train-set characteristics and, thus, there could be a small domain shift in the test-set data. As a result, good test-set results can only be achieved with a robust model and it seems that the utilization of additional datasets and the use of pseudo-labels both increase the robustness of the model significantly.

In table \ref{table:pretrainings} it is clearly visible that all pretrainings yield significantly better results than a randomly initialized model in terms of the cross-validation metrics and the \emph{segmentation}, \emph{segmia} and \emph{multitask} models score is higher than the score of the \emph{ImageNet model}. Consequently, it can be concluded that a pretraining utilizing segmentation labels is highly favorable. As the \emph{multitask model} outperforms the other variants on the test-set as well as on the cross-validation metrics, we think that a pretraining with a task similar to the final task as well as the utilization of segmentation pseudo-labels is very beneficial. We suppose that the models gain a greater robustness due to the usage of the additional datasets in the pretraining pipeline. As the STOIC dataset used for the multitask pretraining is comparably large, the \emph{multitask model} seems to be especially robust, thus performing best on the test-set. Subsequently, we recommend combining a task similar to the final task with a segmentation task for superior pretraining results.

\subsection{Challenge Submission Results}
\label{section:challengesubmission}
We participate in two challenges hosted in the context of the Medical Image Analysis (MIA) workshop at ECCV 2022 (\cite{kollias2022ai}). The two tasks were the detection of COVID infections and the prediction of the severity the patient is experiencing. We rank 2nd in the \emph{1st COVID19 Severity Detection Challenge} and 3rd in the \emph{2nd COVID19 Detection Challenge}. In this section, we present our submissions and further discuss the results. Our submission code is published at \url{https://github.com/KieDani/Submission_2nd_Covid19_Competition}.

%When inspecting the dataset, we found that few CT scans had different orientations. In order to gain robustness with respect to such variations, an augmentation that randomizes the orientation of the CT tensor is implemented and we apply this augmentation with a 25\% probability. If this augmentation is used, it is denoted as \emph{ROr}.
Since we apply 5-fold cross-validation, we suggest to use an ensemble of the 5 trained models to generate the validation-set as well as test-set predictions. However, due to a coding mistake, we only use the fifth model instead of the full ensemble during the challenge. As a result, this causes deviations from the cross-validation estimate as our model is only trained with 80\% of the training data. Nevertheless, it is even more impressive that we still achieved such good results and, thus, our architecture and pretraining pipeline are very well suited for COVID-related tasks.

In contrast to using the fifth model for predictions, we advise to use either an ensemble of the 5 models or to train a single model with all available data based on the settings found with cross validation.

\subsubsection{1st COVID19 Severity Detection Challenge:}
The ranking for the winning teams is shown in table \ref{table:comparisonSeverity}. It can be seen that our method is by a large margin the best in predicting \emph{Severe} and \emph{Critical} cases. We suppose this is achieved through the utilization of the balanced cross entropy and we emphasize that this property is exceptionally important for clinical use cases. \\
\begin{table}[ht]
\caption{Comparison of of best submissions of the winning teams in the 1st COVID19 Severity Detection Challenge. Performance is evaluated for the severity-prediction task. Our prediction is calculated only with the 5th model of the 5-fold cross-validation. F1 scores are macro F1 scores.}
\begin{center}
\begin{tabular}{p{19mm} | p{19mm} | p{19mm} | p{19mm} | p{19mm} | p{19mm}} 
\hline \textbf{Team} & \textbf{F1 Test} & \textbf{F1 Test \newline Mild} & \textbf{F1 Test \newline Moderate} & \textbf{F1 Test \newline Severe} & \textbf{F1 Test \newline Critical} \\
\hline
\textit{1st:} FDVTS & \textbf{51.76} & 58.97 & \textbf{44.53} & 58.89 & 44.64 \\
\textit{2nd}: Ours & 51.48 & \textbf{61.14} & 34.06 & \textbf{61.91} & \textbf{48.83} \\
\textit{3rd}:\newline CNR-IEMN & \hspace{10mm}\newline 47.11 & \hspace{10mm}\newline 55.67 & \hspace{10mm}\newline 37.88 & \hspace{10mm}\newline 55.46 & \hspace{10mm}\newline 39.46 \\
\hline
\end{tabular}  
\end{center}
\label{table:comparisonSeverity}
\end{table}
Since it was possible to submit up to 5 different solutions to the challenge, we list our submissions in table \ref{table:submissionsSeverity}.
\begin{table}[ht]
\caption{Our submissions to the 1st COVID19 Severity Detection Challenge. The ensemble predictions are marked with a \textdagger . Predictions marked with \fifth are calculated only with the 5th model of the 5-fold cross-validation. Usage of the random-orientation augmentation is denoted with \emph{ROr}. F1 scores are macro F1 scores.}
\begin{center}
\begin{tabular}{c |c | c | c | c} 
\hline \textbf{Submission \#} & \textbf{Pretraining} & \textbf{F1 Cross Validation} & \textbf{F1 Test} & \textbf{F1 Validation} \\ 
\hline
% FinalSeverityCV_balce-imagenet_lrDecayUntil0_initModefull_rotProb0.0_AugModeall_cv_convnext_20220621-115412
1 & ImageNet (Full) & 65.67 & 46.67\textsuperscript{\fifth} & 67.21\textsuperscript{\fifth} 61.28\textsuperscript{\textdagger} \\
% FinalSeverityCV_balce-segmentationECCVFull_lrDecayUntil0_initModefull_rotProb0.0_AugModeall_cv_convnext_20220624-183130
2 & Segmentation & 67.25 & 49.36\textsuperscript{\fifth} & 63.43\textsuperscript{\fifth} 61.28\textsuperscript{\textdagger} \\
% FinalSeverityCV_balce-segmiaECCVFull_lrDecayUntil0_initModefull_rotProb0.0_AugModeall_cv_convnext_20220625-180946
3 & Segmia & 66.48 & \textbf{51.48\textsuperscript{\fifth}} & 60.89\textsuperscript{\fifth} 63.05\textsuperscript{\textdagger} \\
% FinalSeverityCV_balce-multitaskECCV_lrDecayUntil0_initModefull_rotProb0.0_AugModeall_cv_convnext_20220621-115510
4 & Multitask & 68.18 & 46.01\textsuperscript{\fifth} & 55.51\textsuperscript{\fifth} 58.77\textsuperscript{\textdagger} \\
% FinalSeverityCV_balce-segmentationECCV_lrDecayUntil0_initModefull_rotProb0.25_AugModeall_cv_convnext_20220625-111222
5 & Segmentation ROr & 71.74 & 49.90\textsuperscript{\fifth} & 60.02\textsuperscript{\fifth} 62.68\textsuperscript{\textdagger} \\
\hline
\end{tabular}  
\end{center}
\label{table:submissionsSeverity}
\end{table}
The \emph{segmia model} performs best. Furthermore, the comparison of submission 2 and 5 indicates that the random-orientation augmentation increases the performance. However, as the test-set scores are very similar the augmentation does not have a great effect on the test set and, thus, is negligible for this challenge. We suppose that this is due to fewer CT-scans with deviating orientation in the test set compared to the train and validation sets. \\

\begin{figure}[t]

\begin{center}
\begin{tabular}{ccc}
    \textbf{Correctly Classified} & \quad & \textbf{Underestimated} \\
    \begin{tabular}{c c}
        \includegraphics[width=0.20\textwidth]{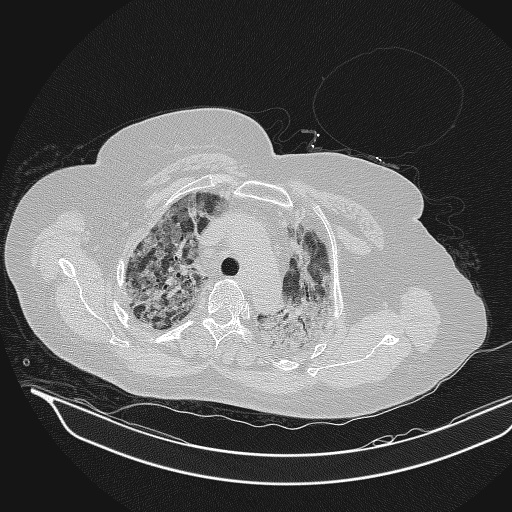} & \includegraphics[width=0.20\textwidth]{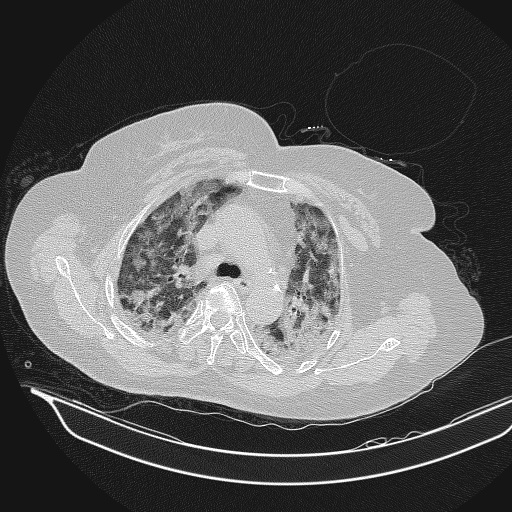} \\
        \includegraphics[width=0.20\textwidth]{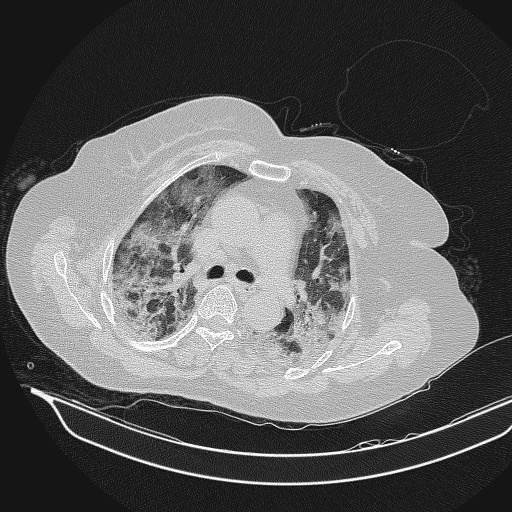} & \includegraphics[width=0.20\textwidth]{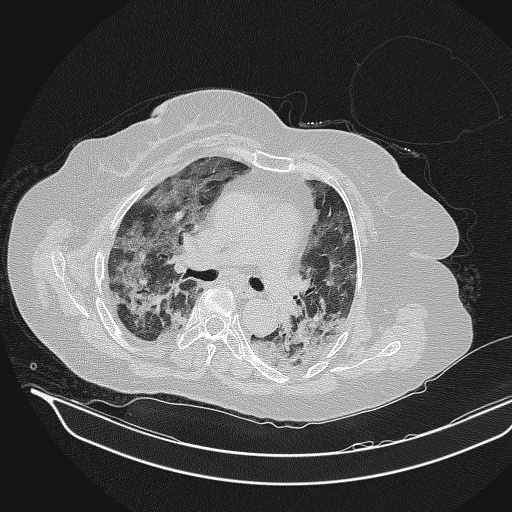} \\
        \includegraphics[width=0.20\textwidth]{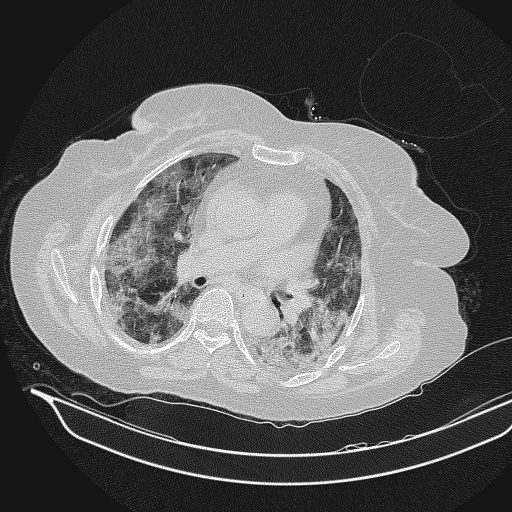} & \includegraphics[width=0.20\textwidth]{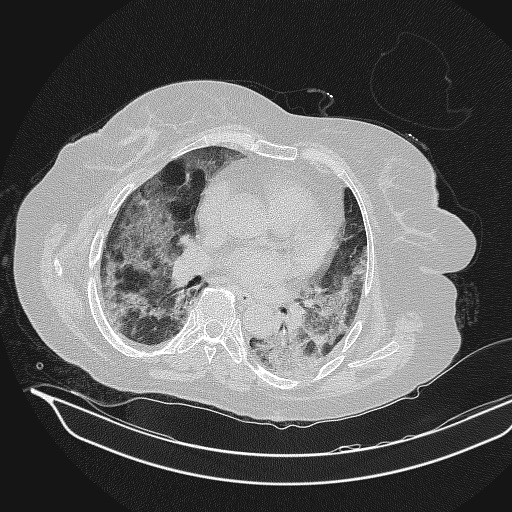} \\
        \includegraphics[width=0.20\textwidth]{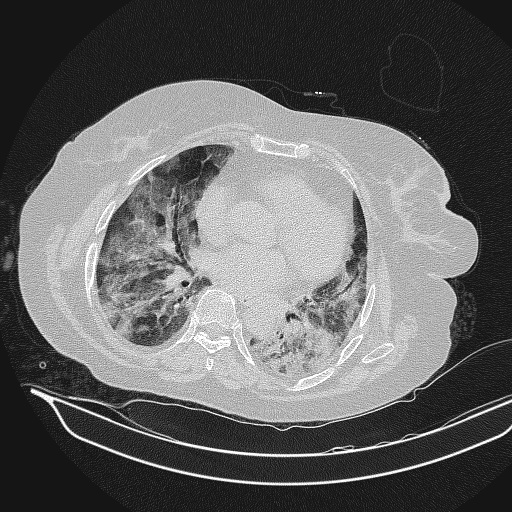} & \includegraphics[width=0.20\textwidth]{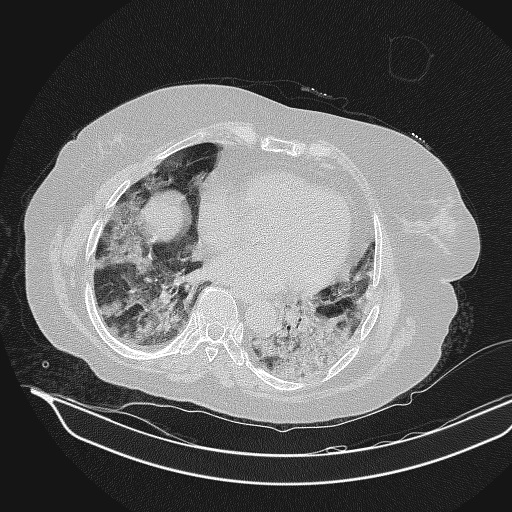} \\
    \end{tabular} 
     & \quad & 
    \begin{tabular}{c c}
        \includegraphics[width=0.20\textwidth]{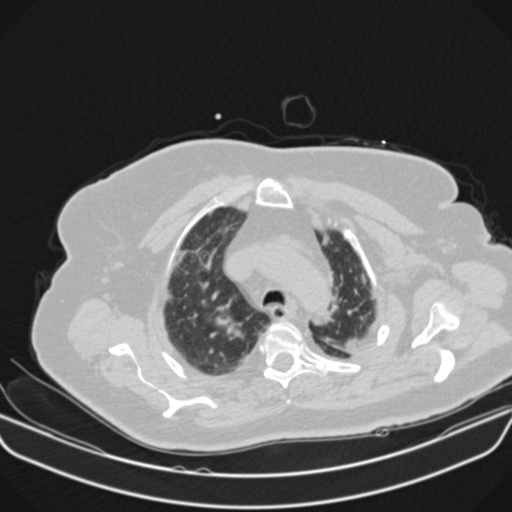} & \includegraphics[width=0.20\textwidth]{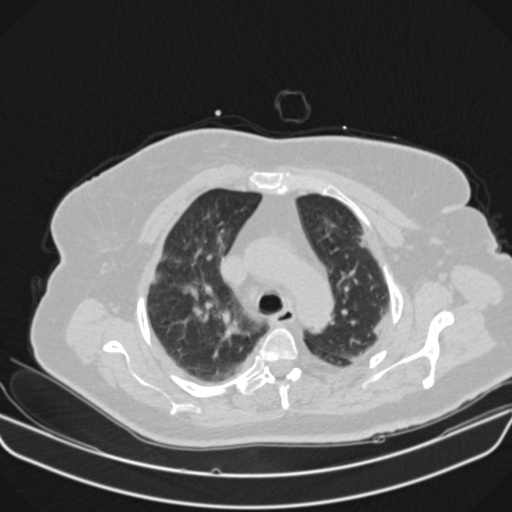} \\
        \includegraphics[width=0.20\textwidth]{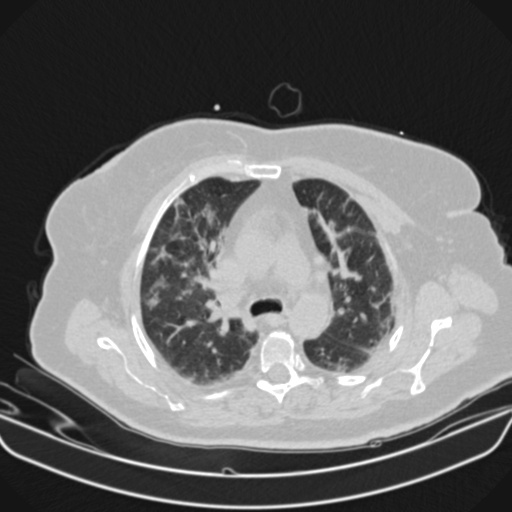} & \includegraphics[width=0.20\textwidth]{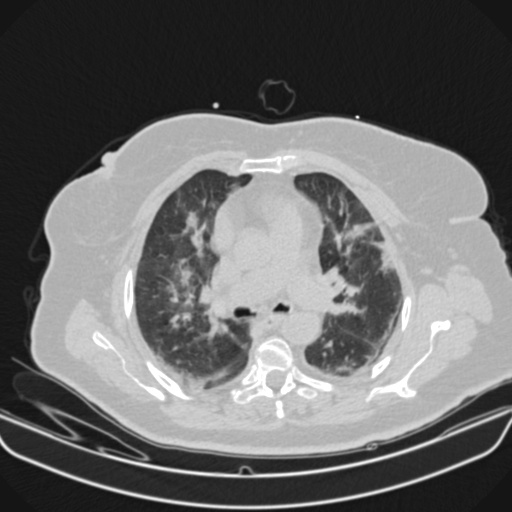} \\
        \includegraphics[width=0.20\textwidth]{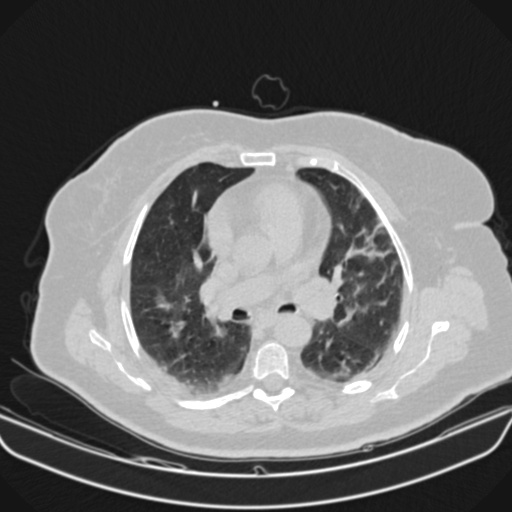} & \includegraphics[width=0.20\textwidth]{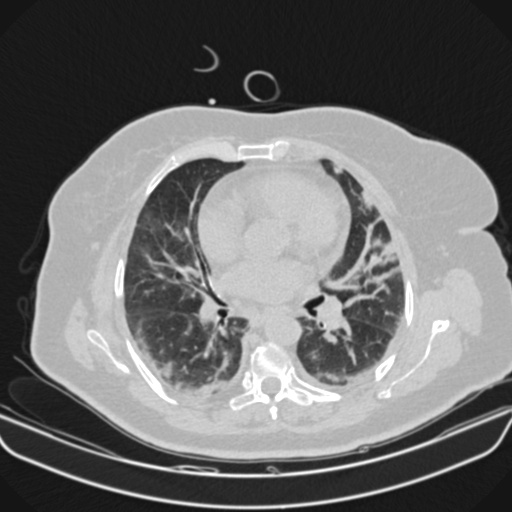} \\
        \includegraphics[width=0.20\textwidth]{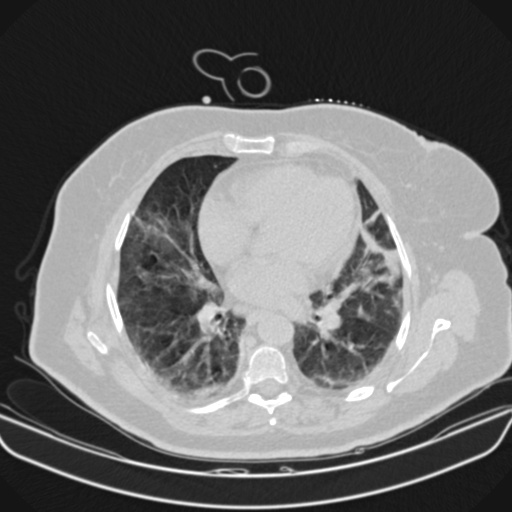} & \includegraphics[width=0.20\textwidth]{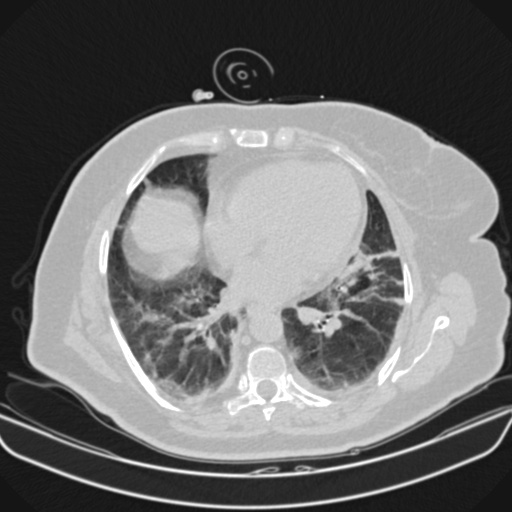} \\
    \end{tabular}
\end{tabular}
\end{center}

\caption{Qualitative example slices for the severity prediction. The CT scan on the \emph{left} has been correctly identified to display a patient that is in a critical state. The CT scan on the \emph{right} has been predicted to be in a moderate state although the correct prediction would have been a severe state.}

\label{fig:exampleslices}

\end{figure}

We added an example for a correctly classified and an incorrectly classified CT-scan in figure \ref{fig:exampleslices}.

\subsubsection{2nd COVID19 Detection Challenge:}
In addition to severity-detection, we used our architecture and training-pipeline also to train our model for the task of infection detection and participated in the \emph{2nd COVID19 Detection challenge}. The ranking for the winning teams is depicted in table \ref{table:comparisonInfection}. 
\begin{table}[ht]
\caption{Comparison of the best submissions of the winning teams in the 2nd COVID19 Detection Challenge. Our prediction is calculated only with the 5th model of the 5-fold cross-validation. F1 scores are macro F1 scores.}
\begin{center}
\begin{tabular}{p{20mm} | p{20mm} | p{23mm} | p{20mm} } 
\hline \textbf{Team} & \textbf{F1 Test} & \textbf{F1 Test \newline Non-COVID} & \textbf{F1 Test \newline COVID} \\
\hline
\textit{1st}: ACVLab & 89.11 & 97.45 & 80.78 \\
\textit{1st}: FDVTS & 89.11 & 97.31 & 80.92 \\
\textit{2nd}: MDAP & 87.87 & 96.95 & 78.80 \\
\textit{3rd}: Ours & 86.18 & 96.37 & 76.00 \\
\hline
\end{tabular}  
\end{center}
\label{table:comparisonInfection}
\end{table}
Because it was also possible to submit up to 5 solutions, our submissions can be seen in table \ref{table:submissionsInfection}.
\begin{table}[ht]
\caption{Our submissions to the 2nd COVID19 Detection Challenge. F1 scores are macro F1 scores. The * denotes that no cross validation was used. The ensemble predictions are marked with a \textdagger .  Predictions marked with \fifth are calculated only with the 5th model of the 5-fold cross-validation. Usage of the random-orientation augmentation is denoted with \emph{ROr}.}
\begin{center}
\begin{tabular}{c |c | c | c | c} 
\hline \textbf{Submission \#} & \textbf{Pretraining} & \textbf{F1 Cross Validation} & \textbf{F1 Test} & \textbf{F1 Validation} \\ 
\hline
% FinalInfectionCV_balce-imagenet_lrDecayUntil0_initModefull_rotProb0.0_AugModeall_cv_convnext_20220621-122204
1 & ImageNet (full) & 91.73 & 82.13\textsuperscript{\fifth} & 86.60\textsuperscript{\fifth} 89.71\textsuperscript{\textdagger} \\
% FinalInfectionCV_balce-multitaskECCV_lrDecayUntil0_initModefull_rotProb0.0_AugModeall_cv_convnext_20220621-125928
% FinalInfectionCV_balce-multitaskECCV_lrDecayUntil0_initModefull_rotProb0.0_AugModeall_cv_convnext_20220621-125928
2 & Multitask & 93.53 & 86.02\textsuperscript{\fifth} & 87.80\textsuperscript{\fifth} 88.79\textsuperscript{\textdagger} \\
% FinalSeverityCV_balce-segmiaECCVFull_lrDecayUntil0_initModefull_rotProb0.0_AugModeall_cv_convnext_20220625-180946
% FinalInfectionCV_balce-segmiaECCV_lrDecayUntil0_initModefull_rotProb0.0_AugModeall_cv_convnext_20220621-125638
3 & Segmia & 93.33 & 83.63\textsuperscript{\fifth} & 88.31\textsuperscript{\fifth} 89.22\textsuperscript{\textdagger} \\
4 & Segmia ROr\textsuperscript{*} & - & 83.93 & 92.03 \\
5 & Segmentation\textsuperscript{*} & - & \textbf{86.18} & 93.48 \\
\hline
\end{tabular}  
\end{center}
\label{table:submissionsInfection}
\end{table}
We achieve the best results without cross validation using the \emph{segmentation model}. This is probably due to the coding mistake mentioned above as this model is trained with 100\% of the training data in contrast to 80\%. Nearly the same performance is achieved using the (fifth) \emph{multitask model}, thus indicating that the multitask pretraining is a good choice for infection detection as well. As submission 2 to 5 are considerably better than submission 1, we conclude that our custom pretrainings improve the results in contrast to the ImageNet model for the infection-detection task, too.

Furthermore, by analyzing submission 1 to 3, we deduce that the cross-validation results are good estimates for the models performance as the order of the scores matches the test-set scores. Moreover, since the gap between cross-validation metrics and test-set metrics is considerably smaller than for the severity prediction task, we reason that the dataset statistics of the train-set and the test-set are much more similar for the infection-detection task. We guess that the statistics are more similar in this challenge because the dataset size is substantially larger and, consequently, we emphasize the need to use larger datasets in order to get valid performance estimates for clinical usage.

\section{Conclusion}
In this paper, we analyzed various pretraining techniques designed to enhance SARS-CoV-2 severity-prediction performance of our neural network and show that the performance can be significantly increased utilizing segmentation labels and additional datasets. Additionally, we show that our architecture and pretraining pipeline can easily  be transferred to the task of infection detection and, thus, our method can be regarded as a general method to enhance COVID-related CT-scan analysis. \\
The pretraining methods were applied to a three-dimensional ConvNeXt architecture and a finetuning for the COV19-CT-DB dataset was performed. We achieved 2nd rank in the \emph{1st COVID19 Severity Detection Challenge} and 3rd rank in the \emph{2nd COVID19 Detection Challenge}, consequently proving that our method yields competitive results. \\ 
In addition to that, we introduced the balanced cross-entropy and argued that this loss-function is important for clinical use cases. We emphasize that our model achieved best results in detecting the most-severe cases. \\
Altogether, we presented a framework for severity prediction as well as infection detection and achieved good performance by applying this framework to the ConvNeXt architecture. We encourage further research based upon our framework to enhance the diagnosis options in clinical use cases.

\bibliographystyle{splncs04}
\bibliography{egbib}

%\end{comment}
\end{document}